\documentclass{article}

\usepackage{amsmath}
\usepackage{amssymb}
\usepackage{balance}
\usepackage{graphicx}
\usepackage{caption} 
\usepackage{color}
\usepackage{blindtext}
\usepackage[frenchb,english]{babel}
\usepackage[none]{hyphenat}
\usepackage{algorithm,algpseudocode}
\usepackage{rotating}
\usepackage{multirow}
\usepackage{array}
\usepackage{booktabs,caption,float,multirow,makecell,tabularx}
\newcolumntype{l}{>{\centering\arraybackslash}p{1cm}<{}}
\newcolumntype{n}{>{\centering\arraybackslash}p{3cm}<{}}
\usepackage{xcolor}

\usepackage{hyperref}
\usepackage{lmodern}
\usepackage{textcomp}
\usepackage[sc]{mathpazo} 
\usepackage[T1]{fontenc} 
\usepackage{microtype} 
\usepackage[hmarginratio=1:1,top=25mm,columnsep=20pt]{geometry} 
\usepackage{float} 
\usepackage{paralist} 
\usepackage{abstract} 
\usepackage{titlesec} 
\renewcommand\thesubsection{\Roman{subsection}} 
\titleformat{\section}[block]{\large\scshape\centering}{\thesection.}{1em}{} 
\titleformat{\subsection}[block]{\large}{\thesubsection.}{1em}{} 
\usepackage{fancyhdr} 
\begin{document}
\title{On Improving Service Chains Survivability\\Through Efficient Backup Provisioning}
\date{}

\author{
Saifeddine Aidi\textsuperscript{\textdagger},
Mohamed Faten Zhani\textsuperscript{\textdagger},
Yehia Elkhatib\textsuperscript{\textdagger}\textsuperscript{\textasteriskcentered}\\[4mm]
\small {\textdagger} \'Ecole de Technologie Sup\'erieure (\'ETS Montreal), Montreal, Quebec, Canada\\
\small {\textasteriskcentered} MetaLab, School of Computing and Communications, Lancaster University, UK\\
E-mail: saifeddine.aidi.1@ens.etsmtl.ca, mfzhani@etsmtl.ca, \{i.lastname\}@lancaster.ac.uk\\
    \textbf{\textcolor{red}{This is a pre-print. Please cite the CNSM version of this paper.}}\\
}

\maketitle

\thispagestyle{fancy}

\begin{abstract}
With the growing adoption of Software Defined Networking (SDN) and Network Function Virtualization (NFV), large-scale NFV infrastructure deployments are gaining momentum. Such~infrastructures are home to thousands of~network Service Function Chains (SFCs), each composed of~a~chain of~virtual network functions (VNFs) that are processing incoming traffic flows. 
Unfortunately, in such environments, the~failure of~a~single node may break down several VNFs and~thereby breaking many service chains at the same time. 

In~this~paper, we address this particular problem and~investigate possible solutions to ensure the~survivability of~the~affected service chains by~provisioning backup VNFs that~can take over in case of~failure. Specifically, we~propose a~survivability management framework to efficiently manage SFCs and the backup VNFs. We~formulate the SFC survivability problem as an integer linear program that determines the~minimum number of required backups to~protect all the~SFCs in the system and identifies their optimal placement in the infrastructure. 
We~also~propose two heuristic algorithms to~cope with the large-scale instances of~the~problem.
Through~extensive simulations of~different deployment scenarios, we~show that~these~algorithms provide near-optimal solutions with minimal computation time.
\end{abstract}

\section{Introduction}

The emergence of Network Function Virtualization (NFV) and Software-Defined Networking (SDN) technologies is~currently transforming the way networks are designed and~managed as they provide operators much more flexibility to~dynamically provision and configure network services. 
In~particular, it is now possible to~dynamically create chains of~network services (Service Function Chains - SFCs) that can process incoming traffic and~steer it across a chain of Virtual Network Functions~(VNFs) like routers, IDSs and~NATs that~are~running on~virtual machines.

In the last few years, a large body of work has been dedicated to address resource provisioning and management of~such~SFCs~\cite{ZHANISurvey2013,HerreraSurvey016, r8, r10, r11, Luizelli2015}. Most of existing studies assume the~complete availability of the physical infrastructure which is not realistic as failures are common in cloud network infrastructures \cite{lee2017overload, ZDNetFailures, ETSI, ZhangVenice2014}. 
Due to the dependency between virtual network functions in the chain, a single physical node failure in the network could easily bring down many VNFs and hence break several SFCs and make these services unavailable. Such downtime, even for few seconds, not only hurts the reputation of service providers but also incurs high revenue losses depending on~the~type of the offered service (e.g.,~\$5,600~ per minute according~to~\cite{OutageCosts}).

Existing proposals to manage failures and mitigate them can be broadly categorized into reactive and proactive techniques \cite{Zhani2015Surv}. In~proactive techniques, backup VNFs are provisioned whenever an SFC is received and embedded. These backups remain idle but are activated only when a failure occurs to take over the~service and replace the failed VNFs  \cite{r3, r4, r5, RabbaniIEICE13}. 
The second category of existing solutions are reactive techniques. These techniques do not pre-allocate backup resources and deal with failures after they occur~\cite{r5,r6, r7}. Consequently, they need additional time to allocate resources and provision new VNF instances to take over the service. This definitely results in a longer service disruption, which is very costly for service providers \cite{Zhani2015Surv}. This makes proactive techniques more appealing even though they waste some resources for~backup~VNFs. 

To remediate to this problem and minimize wastage of~resources, several research efforts advocate to use shared backup VNFs \cite{r4} where the same backup resource can be used to mitigate the failure of a set of  VNFs assuming that they do not fail at the same time (i.e.,~only a single VNF from this~set can fail at a time).
In this context, this~paper investigates possible solutions to ensure the survivability of~service chains against single physical node failures by using shared backup resources. Unlike previous work addressing the~same problem where backups are shared only between the~VNFs of the same chain \cite{r3} \cite{r4} , our solution assumes that backup VNFs are shared among all the chains embedded in~the~infrastructure. This significantly reduces the amount of~resources used for~the~backup VNFs while still ensuring all SFCs are protected against single failures. 

\begin{figure*}[t!]
	\centering
	\includegraphics[scale=0.6]{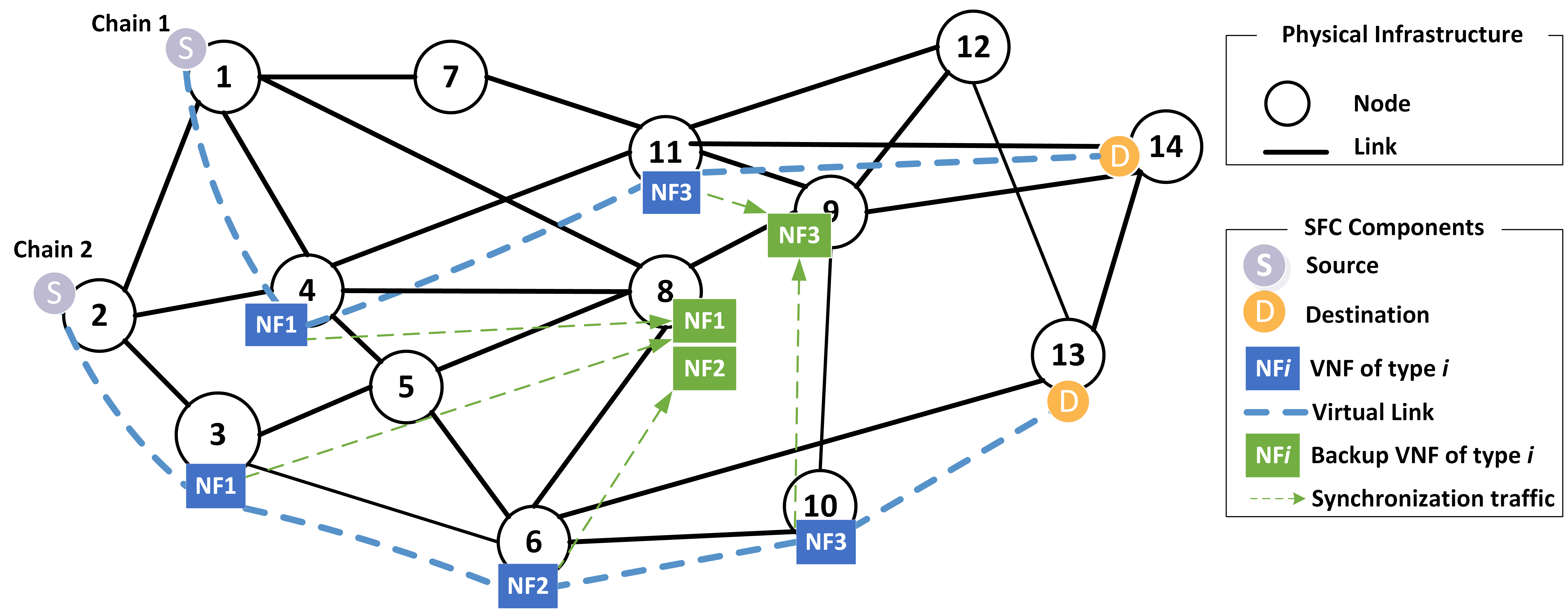}
	\caption{An example of various embedded SFCs sharing backups.}
	\label{fig:ProblemDescription}
\end{figure*}

Our main goal is to ensure the survivability of~all~embedded SFCs against any single-node failure in~the~physical infrastructure. We reach this objective by proactively provisioning the minimal number backup VNFs to~minimize resource wastage and by carefully placing them in~the~infrastructure. We also take into consideration the~synchronization cost in~terms~of~bandwidth and~delay needed to keep the backup nodes up-to-date. 

We~can~summarize the~main contributions of this paper as~follows:
\begin{itemize}
	\item We propose a~resource management framework with~a~survivability module. This module allows to~provision and manage backup VNFs and~it~could~be easily integrated into existing SFC resource management~frameworks.
	\item We formulate the backup provisioning and placement problem as an Integer Linear Program (ILP) that finds the~optimal number of shared backups for each type of VNF and optimally places them in the physical infrastructure. 
	\item We devise two heuristic algorithms, called BS-Push and~BS-Pull, respectively, that aim at solving the problem for~large-scale scenarios within a reasonable timescale.
	\item We evaluate the performance of the proposed heuristics and compare it to the optimal solution found with~the~proposed ILP solved by~the~CPLEX optimizer. 
\end{itemize} 

The remainder of this paper is organized as follows. Section~\ref{Section:ProbDescription} provides a detailed description of service chain survivability problem. We discuss relevant related work in~Section~\ref{Section:RelatedWork}. In~Section~\ref{Section:Framework} and~\ref{Section:Solutions}, we mathematically formulate the~addressed problem and then describe the~proposed heuristic solutions. We present the experimental results in Section~\ref{Section:Simulations} and follow up with some conclusions and future work in~Section~\ref{Section:Conclusion}.

\section{Problem Description}\label{Section:ProbDescription}

A Service Function Chain is made out from a set of different types of virtual network functions connected in a specific order to form a chain that steers the traffic from and to predefined source and destination~\cite{r8}. A Virtual Network Function (VNF) is simply a virtual resource (i.e., virtual machine or container) that is running a specific network function (e.g.,~router, load balancer, NAT, IDS). 
To build the chain, the~VNFs are connected through a~set of virtual links having a sufficient amount of bandwidth to handle the traffic. 
Typically, service function chains are embedded into a physical infrastructure (referred~to~as NFV Infrastructure~-~NFVI~\cite{ETSI}). 

Figure~\ref{fig:ProblemDescription} shows~an~example of two service chains mapped onto a wide area NFV infrastructure. The figure shows for~each chain how the VNFs are embedded from the source to~the~destination. For~instance, chain 2 has traffic coming from physical node 2 towards physical node 13 and~it~is~composed of three VNFs of type 1, 2 and 3 that~are~embedded in~physical nodes 3, 6 and 10, respectively. Chain 1 has only two VNFs of type 1 and 3 that are embedded in physical nodes~4~and~11,~respectively.

Once SFCs are embedded into the NFVI, the operator faces the challenging task of ensuring the high survivability of~these SFCs. In other words, they need to survive potential network failures in order to minimize service interruptions. However, as mentioned earlier, physical nodes are prone to~failures and~a~single node failure may result in bringing down several VNFs and~hence breaking multiple service chains. In~this~paper, we propose to ensure the survivability of the SFCs affected by a single failure by leveraging shared backups that could be used when the failure occurs. We~also~assume that a backup VNF should be of~the~same type of~the~set~of~VNFs it is backing up. In other words, a VNF of~type~$i$ can only back up VNFs of type $i$. This~assumption is reasonable as in practice the backup is~a~virtual machine that should implement a specific software and~hence~a~backup VNF has~to~contain exactly the same software stack as~the~original~VNF.

As an example of how shared backups could be placed, we~can see in Figure~\ref{fig:ProblemDescription} that physical node 9 hosts a backup of VNF type 3 (i.e., NF3) that is shared between the VNF type~3 of~chain~1 and~that~of~chain 2. If physical node~11 fails, and~hence~NF3 of~chain 1 becomes out of service, the backup NF3 hosted in 9 takes over and replaces the~failed function. Similarly, it can take over the service of NF3 belonging to~chain~2 (hosted in node 10) if it fails. The figure also shows other examples of shared VNF backups (e.g., NF1 and NF2 hosted in node 8).
It is easy to check that the two service chains shown in this example are perfectly survivable to~any~single node failure.

Furthermore, backup VNFs are continuously synchronized with the active VNFs to be ready to take over the service in~case~of~failure (see green arrows in Figure~\ref{fig:ProblemDescription}). For instance, the backup NF3 hosted in 9 has the state of NF3 hosted in~physical node 11 and that of NF3 hosted in~10. Whenever a failure happens, the last state of the failed function will~be used when the backup is activated. State synchronization can be done at a different level. For example, at the level of the virtual machine running the function (e.g., memory synchronization~\cite{ZHANI-IGI-Global13}) or~using customized synchronization scripts depending on~the~type of~the~network function (e.g.,~synchronizing rules in~firewalls). 

To make sure that state synchronization is efficient, the~latency between a VNF and its backup should not exceed a~certain bound. Furthermore, synchronization of the VNF state may consume bandwidth that should be minimized. In~our~work, we ensure to minimize the synchronization delay and~consumed bandwidth by limiting the number of hops between each VNF and its backup (for example, the number of hops is limited to 2 in ~Figure~\ref{fig:ProblemDescription}).

The main challenge that we are addressing in this paper is~how to find the minimal number of backup nodes and to determine their optimal placement in the physical infrastructure for each type of~VNF taking into account the~synchronization delay and the cost in terms of bandwidth consumption. 

\section{Related Work}\label{Section:RelatedWork}

In this section, we provide an overview of representative work on the survivability problem. We note that most existing work focuses on virtual networks survivability and not SFCs. However, they are still valid for our case as a service function chain can be seen as a particular case of a virtual network with a specific topology. In the following, we summarize existing techniques to ensure the survivability of SFCs and virtual networks as either \emph{reactive} or \emph{proactive}~\cite{Zhani2015Surv}. 

Reactive techniques do~not~pre-allocate resources for~backup but simply deal with a~failure when it occurs. This leads to~a~long convergence time after the failure, resulting in~a~higher service downtime. On the other hand, proactive solutions anticipate failures and pre-allocate backup resources to ensure fast recovery of the service in case of failures. 

Yu~et~al.~\cite{r3} considered the case of a single-node failure and introduced two approaches to provision backup nodes. The first approach,  called 1-redundant, redesigns the virtual network request into a survivable request by~adding a~single backup node. The second approach is called $k$-redundant where $k$ is a constant that represents the~number of backup nodes to be provisioned. 
The problem with these approaches is that a single redundant node may not be enough whereas $k$ redundant nodes might be too much, and hence could lead to~a~wastage of~resources. To~address this limitation, the~solutions presented in this current work aim at finding the~optimal number of backup nodes when~the~proposed ILP is used or at least minimize it~when the~proposed~heuristics~are~used.

In the same direction, Ayoubi et al. \cite{r4} explored the space between $1$ and $k$ to find the optimal number of backup nodes to be incorporated into the requested virtual network. However, in this solution, the backup virtual nodes are provisioned for each request and hence they are not shared with other virtual networks. Our~work is different in that it provisions backups that are shared among all virtual nodes belonging to~all~virtual infrastructures (i.e., SFCs) embedded in~the~physical infrastructure. As a result, our solutions further reduce the total number of backups provisioned in the system. 

\begin{sidewaystable}
	\centering
	\caption{Existing solutions vs. the proposed ones\label{tab:servact}}
	\begin{tabular}{|c|c|c|c|c|c|c|c|c|c|}
		\hline
		\multirow{3}{2cm}{\centering \bf Solutions } & \multicolumn{2}{|c|}{\bf Type of solution} & \multicolumn{2}{|n|}{\bf \makecell{Single/Multiple \\ failures}} & \multicolumn{2}{|c|}{\bf \makecell{Node/Link \\ failures}} & \multicolumn{3}{|c|}{\bf \makecell{Support of Shared backups}} \\
		\cline{2-10} 
		& Proactive & Reactive & Single & Multiple & Node & Link & Supported & \makecell{Shared among \\all VNs} & \makecell{Shared among\\ a~single VN} \\
		\hline
		Yu et al. \cite{r3} & x & & x & & x & & x & & x \\
		\hline
		Ayoubi et al. \cite{r4} & x & & x & & x & & x & & x\\
		\hline
		Guo et al. \cite{r9} & x & & x & & x & & & - & - \\
		\hline
		Xiao et al. \cite{r6} & x &  & & x & x & & & - & - \\
		\hline
		Rahman et al. \cite{r5} & x &  & x & & & x & & - & - \\
		\hline
		Bo et al. \cite{r7} & & x & x & & x & & & - & - \\
		\hline 
		Ayoubi et al. \cite{r13} & & x & x & & x & & & - & - \\
	  \hline 
		Ghaleb et al. \cite{r12} & & x & & x & x & x & & - & - \\
		\hline
		BS-Pull/BS-Push & x & & x & & x & & x & x & \\
		\hline	
	\end{tabular}
\end{sidewaystable}

 Xiao~et~al.~\cite{r6} proposed a topology-aware solution that ensures a rational resource allocation for the virtual network and a fail-over remapping based on a set of pre-computed detour-paths. Rahman~et~al.~\cite{r5} also proposed a hybrid approach that~benefits from a set of a possible backup detours for each link. These detours are proactively precomputed before the arrival of virtual network requests to allow fast re-routing in case of link failure. 

Finally, Bo et al. \cite{r7} proposed a greedy algorithm that, in case of~a~link failure, searches for alternative resources to re-allocate the end-to-end path or re-embed the entire virtual network if resources are not sufficient. This~may result in~long convergence time and~higher service downtime. 
In~\cite{r13}, Ayoubi et al. have demonstrated the~NP-hard nature of~the~survivability-aware embedding
and~proposed a polynomial time heuristic algorithm to~restore failed services while maintaining the QoS requirements in~terms of delays in case of~single-node failures. Multiple failures were addressed~in~\cite{r12} where a heuristic was introduced in order to find a~backup node. 
The algorithm is based on filtering techniques to~parse the~solution space and~to~speed~up~the~search process for~backups.

We summarize in Table~\ref{tab:servact} the~aforementioned solutions. The~table presents the type of solution (i.e.,~reactive vs.~proactive) and it indicates whether it is addressing a~single or multiple failures, node or link failures and whether the~backup are shared between the virtual nodes of all virtual networks or among the virtual nodes of~a~single virtual network.   
As shown in the Table, the~novelty of~our~work lies in the idea of~sharing backups between VNFs of~the~same type that belong to~different~or virtual networks (or~service~chains) rather than the same virtual network (or~service~chain), which further reduces the amount of~backup resources while still ensuring the survivability of~the~virtual networks to~any~single~failure.

\section{Survivability Management Framework}\label{Section:Framework}
\begin{figure}[h]
	\centering
	\includegraphics[scale=0.5]{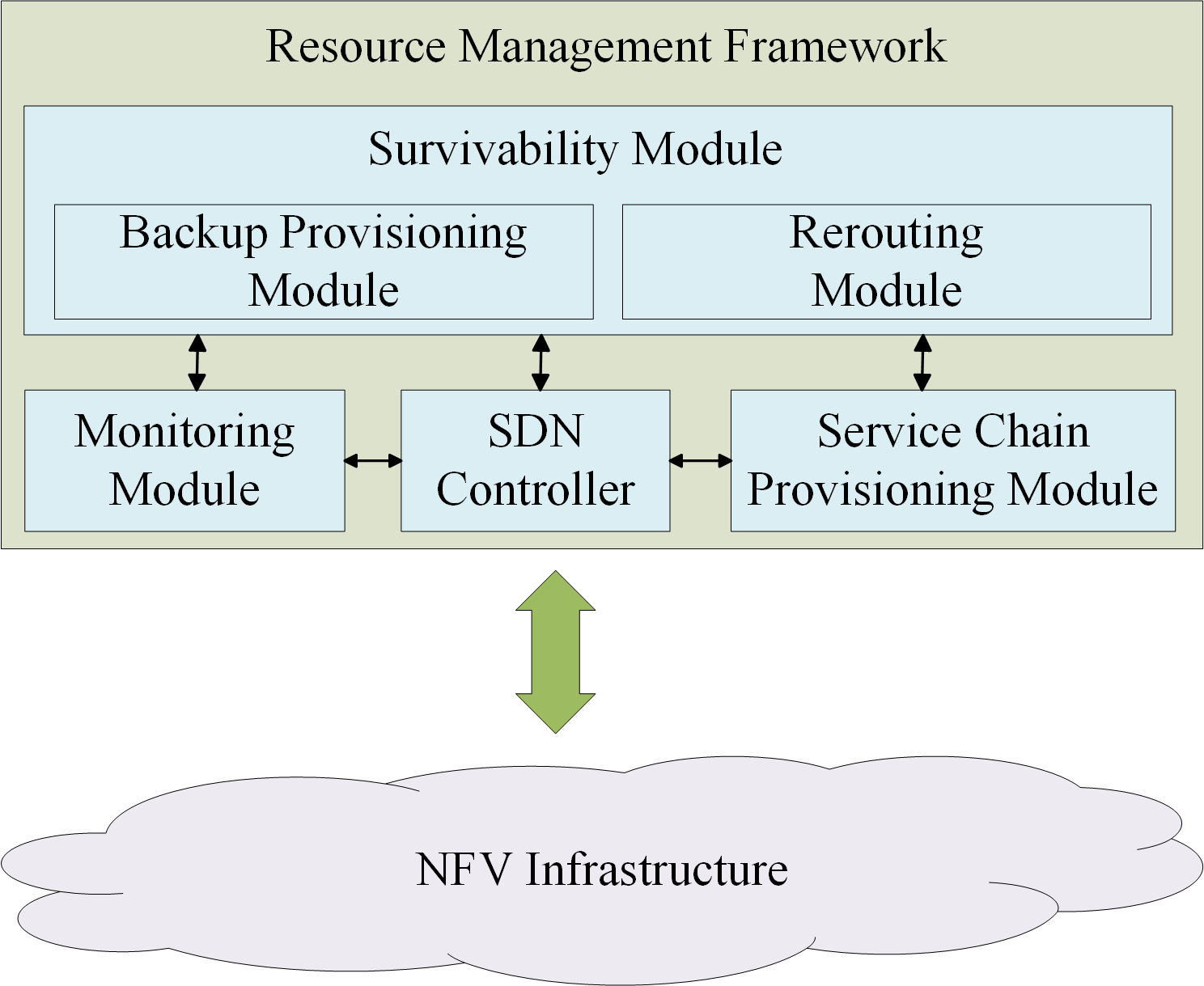}
	\caption{Architecture of the Proposed Resource Management Framework with the Survivability Module.}
	\label{fig:framework}
\end{figure}

In this section, we propose a management framework that incorporates a survivability module. 
Figure \ref{fig:framework} shows the~main components of this framework. It is made out from the~following modules: \\

\textbf{$\bullet$ Service Chain Provisioning Module:} This module allocates the resources for the service chains and~instantiates the required virtual machines running the network functions. It~also~makes use of the SDN controller to~provision the~required amount of bandwidth and~to~set~the~required forwarding rules into the switches to steer the~traffic across~the~VNFs composing each service chain. It~is~worth~noting that~the~design of this module is out of the scope of~this~work. There is a large body of work addressing this module and~any~of~the~existing solutions could be~used~(e.g.,~\cite{r10,Luizelli2015, HerreraSurvey016}).\\

\textbf{$\bullet$ Monitoring Module:} This module is in charge of~continuously monitoring the infrastructure physical nodes and~links and of feeding other modules at real time with the state of~the resources. When a failure is detected, the monitoring module reports to the survivability module, which, in turn,  reacts to~mitigate the failure and ensure service continuity.\\

\textbf{$\bullet$ Rerouting Module:} In case of failure, the rerouting module redirects the traffic, that is originally destined to~the~failed VNFs, to~the backup VNFs.\\

\textbf{$\bullet$ Backup Provisioning Module:} This module is responsible for finding the minimal number of backups needed to ensure the survivability of the embedded chains and to determine~their~locations. This module also instantiates the backup VNFs and~the~synchronization links required to keep them up-to-date. In the following section, we describe in details the~proposed solutions to provision backup VNFs while achieving the~sought-after objectives. 

\section{Backup Provisioning solutions}\label{Section:Solutions}

\subsection{Integer Linear Program}
In this section, we formulate the service chain survivability problem as an ILP aiming at minimizing the~amount of~resources allocated for the backup instances while ensuring a~minimal synchronization cost and delay. \\

\textbf{$\bullet$ Infrastructure and chain modeling:} We model the physical infrastructure as a graph denoted by~$G = (N, L)$ where $N$ is the set of~physical nodes and $L$ is the set of~physical links connecting them. Each physical node $n \in N$ has a computing capacity \textit{c}$_{n}$.
The capacity $c_{n}$ is expressed as the maximal number of VNFs that can be hosted by physical node $n \in  N$. For sake of simplicity, we assume that each VNF is running on a single virtual machine with a standard size. We hence can see $c_{n}$ as the maximal number of virtual machines that could be provisioned in the physical node $n$.
We~also define $d_{in}$ as the minimum number of hops separating the~physical nodes~$i$~and~$n$.

We model the service chain as a graph denoted by~$S = (V,E)$ where $V$ is the set of its composing VNFs and $E$ is the set of virtual links connecting them. We~assume there are different types of VNFs (e.g.,~Firewall, IDS, NAT). We denote by $J$ the set of VNF types and~we~define \textit{m}$_{ij}$ as~the number of VNFs of type $j\in J$ embedded in~physical node $i$.\\

\textbf{$\bullet$ Decision Variables:} We define two decision variables. We denote by \textit{x}$_{ij}$ the number of backup VNFs of type $j$ embedded in physical node $i$. We also define \textit{y}$_{ijn}$ $ \in $ \{0,1\} to indicate whether the backups provisioned for~the~VNFs of type $j$ embedded in the physical node $i$ are hosted in~the~physical node $n$. \\

\textbf{$\bullet$ Problem constraints:} In order to find a feasible solution, several constraints must be satisfied. For instance, to~ensure that the primary and backup VNFs are not embedded in~the~same physical node, the following constraint must~be~satisfied:
\begin{equation}
y_{iji} = 0\quad	\forall i \in N, \forall j \in J
\end{equation}

We also need to ensure that all VNFs of type $j$ embedded in the physical node $i$ necessarily have backups in another physical node:
\begin{equation}
\sum_{n\in N} y_{ijn}  = 1\quad \forall i \in N, \forall j \in J
\end{equation}

Furthermore, if the backups for the type $j$ VNFs embedded in the physical node $i$ are hosted in physical node $n$ then the total number VNF backups of type $j$ provisioned in $n$ should be higher or equal than the number of VNFs of type $j$ embedded in node $i$. In other words, we have:
\begin{equation}
if\quad y_{ijn} = 1\quad then\quad x_{nj} \geq m_{ij}\quad \forall i,n \in N, \forall j \in J
\end{equation}
This if statement can be translated as the following constraint:
\begin{equation}
m_{ij} \leq x_{nj} + M(1 - y_{ijn})\quad \forall i,n \in N, \forall j \in J
\end{equation}
where $M$ is a constant with a large value (in the vicinity of~$10,000$).

Furthermore, to ensure that the physical node hosting the~backups have sufficient resources, the following capacity constraint must be satisfied for every physical node $n$:
\begin{equation}
\sum_{j\in J} m_{nj} + \sum_{j\in J} x_{nj} \leq c_{n}\quad \forall n \in N
\end{equation} 
where the first term represents the number of VNFs hosted in~ the physical node $n$ and the second term represents the~number of VNF backups hosted in the same physical node.

Finally, as we have to~minimize the synchronization cost and delay between the VNFs and their backups, we limit the~number of~hops between each VNF and its backup to~a~limited number of hops denoted by $d_{max}$. Thus,~we~have:
\begin{equation}
if\quad y_{ijn} = 1\quad then\quad d_{in} \leq  d_{max} \quad \forall i,n \in N,\ \forall j \in J   
\end{equation} 
The previous statement can be also written as the following constraint:
\begin{equation}
d_{in} \leq d_{max} + M(1 - y_{ijn})\quad \forall i,n \in N, \forall j \in J
\end{equation}
where $M$ is a constant with a large value.

It is also worth noting that, for sake of simplicity, we~assume that the number of hops between two physical nodes reflects the time delay between them. However,~this~may not~be~always true. In this case, our model can be easily updated to consider the propagation delay between the nodes by~defining $d_{in}$ as the delay of the shortest path between nodes $i$ and $n$, and $d_{max}$ as the maximum delay required between a VNF and its backup.\\

\textbf{$\bullet$ Objective function:} Our ultimate goal is~to~minimize the~amount of resources used by~the~backup VNFs while~satisfying all the~aforementioned constraints. This~can~be~achieved by~minimizing the total number of backups in~all the physical nodes of~the~physical infrastructure. The objective function can be~then written as:
\begin{equation}
\min \sum_{i\in N} \sum_{j \in J} x_{ij} 
\end{equation}

\subsection{Heuristic Algorithms}

\let\oldComment\Comment
\renewcommand{\Comment}[1]{\textit{\oldComment{#1}}}

\begin{algorithm}[!b]
\caption{BS-Pull}\label{BS-Pull}
\begin{algorithmic}[1]
\State \textbf{Inputs}
\State $N$: set of physical nodes
\State $u_{n}$: total number of VNFs hosted in physical node $n$
\State $m_{ij}$: number of type $j$ VNFs hosted in physical node $i$
	
		\For{$j \in J $}  \Comment{Parsing VNF types $\qquad\qquad\qquad$}
		\State $BNodes(j)\gets \O$\Comment{set of physical nodes whose $\ \ \ \ $\newline \hspace*{9em}  type $j$ VNFs have already backups} 
				\Repeat
				\State $\!\!\!\!$\Comment{Parsing all potential hosting nodes$\qquad\qquad\quad$}
					\For{$n \in N$} 
						 	\State $SNeigh(n)\!\gets \O\!$ \Comment{set of source neighbors \newline \hspace*{13.5em} for physical node $n$}
							\State $b_n\!\gets 0\!\!$ \Comment{number of type $j$ VNFs able to \newline \hspace*{10em} use shared VNF backups hosted \newline \hspace*{10em} in node $n$}
					    \State $\!\!$\Comment{Finding source neighbors of $n\qquad\qquad\qquad$}
							\For{$i \in N \backslash (BNodes (j) \cup \{n\}) $} 
									\If {$\!d_{ni}  \leq  d_{max}\And m_{ij} + u_n  \leq c_{n}$}
									\State $SNeigh(n) \gets SNeigh(n) \cup \{i\} $ 
									\State $b_n \gets b_n + m_{ij} $
									\EndIf
							\EndFor
					\EndFor				
						   
					\State $\!\!\!$\Comment{Finding the node $n_{host}$ that maximizes the $\qquad$ \newline \hspace*{3.3em} number of type $j$ VNFs that are backed up$\ \ \ $}
					\State $n_{host}=\operatorname{arg\,max}_n b_n$
					\State $\!\!\!$ \Comment{Compute the number of shared backups $s$$\qquad\ \ $}
					\State $s=\max_{i\in SNeigh(n_{host})}(m_{ij})$
					\State $\!\!\!\!$ \Comment{Allocate backups and update $u_{n_{host}}$ $\qquad\qquad\ \ $}
				  \State Allocate ($s$ backups, VNF type $j$, host $n_{host}$) 
					\State $BNodes(j) \gets BNodes(j) \cup SNeigh(n_{host})$
				\Until{$SNeigh(n)=\O\ \ \forall n \in N$}
		\EndFor
\end{algorithmic}
\end{algorithm}
In this section, we will present two heuristic solutions designed to solve the~survivability problem. We call the~first algorithm \textit{Backup Sharing ``Pull"} (BS-Pull) as~we~are~looking at each physical node to find the maximum number of~VNFs that it~can~backup (we~refer to~this~as~\emph{pulling}). The~second~algorithm is called \textit{Backup Sharing ``Push"} (BS-Push) as~the algorithm tries to push the coverage of~the~physical node in order to let it host backups of VNFs that~are~as~spread as possible in several physical nodes.

Both algorithms are carried out in two phases. The first phase aims at finding the candidate physical nodes that satisfy the constraints of number of hops and the capacity for~each type of VNFs. The second phase aims at selecting among the candidate nodes the ones that should host the backups. The~difference between the two algorithms lies in the way the~candidates hosting nodes are selected.
In the following, we provide the details of~the~two~proposed algorithms.\\
\textbf{$\bullet$ Algorithm BS-Pull:} Algorithm 1 describes the BS-Pull algorithm. 
It aims to allocate VNF backups for each VNF type one by one. Assuming we consider VNF type $j\in J$ first, all nodes in the physical infrastructure are assumed to be able to host backups for type $j$ VNFs. Our goal in~the~following steps is to select which node or nodes could really host these~backups and how many backups per~node. 

We first define \emph{the source neighbors} of~a~physical node~$n$ (i.e.,~$SNeigh(n)$) as~the~set of~physical nodes that could be reached from $n$ within at~most~$d_{max}$ hops and~such~as~node~$n$ has enough resources to~host the backup VNFs required to~back up type $j$ VNFs hosted in any of~these~source neighbors.
In~other~words, if~the~backup VNFs are provisioned in~node~$n$, they can be shared among all the source neighbors~of~$n$. 

For each physical node $n \in N$, we compute the set of~source neighbors~$SNeigh(n)$ and~we~also~compute $b_n$, which~is~the~number of VNFs of type $j$ that could share the~VNF~backups that could be provisioned in~physical node~$n$ (Lines~9-19). The higher $b_n$ is, the higher is~the~number of VNFs sharing the backups. As a result, to~maximize backup sharing, we select the node $n_{host}\in N$ that~has~the~highest value of $b_n$ to~be~the~hosting node of~the~VNF backups. We then allocate the backup VNFs in~the~node~$n_{host}$ (function~Allocate~in~Line~25). 
We repeat this operation until no source neighbors could be identified for~all~the~physical nodes. Having no source neighbors for all physical nodes means that either there is no enough resources to~host the~backup VNFs (while satisfying the constraint on~the~number of~hops) or~there~is~no~VNFs~of~type $j$ that~are~left without backups. 
Finally, the whole process is~repeated for~all~VNF~types.\\

\begin{algorithm}[!h]
\caption{BS-Push}\label{BS-Push}
\begin{algorithmic}[1]
\State \textbf{Inputs}
\State $N$: set of physical nodes
\State $u_{n}$: total number of VNFs hosted in physical node $n$
\State $m_{ij}$: number of type $j$ VNFs hosted in physical node $i$
	
		\For{$j \in J $}  \Comment{Parsing VNF types $\qquad\qquad\qquad$}
		\State $BNodes(j)\gets \O$\Comment{set of physical nodes whose $\ \ \ \ $\newline \hspace*{9em}  type $j$ VNFs have already backups} 
				\Repeat
							\State $\!\!\!$Compute $SNeigh(n)\ \ \ \forall n \in N$ 	
					\State $\!\!\!$\Comment{Finding the node $n_{host}$ that is connected to$\qquad$ \newline \hspace*{3.3em} to the maximum number of neighbors$\ \ \ $}
					\State $n_{host}=\operatorname{arg\,max}_n |SNeigh(n)|$
					\State $\!\!\!$ \Comment{Compute the number of shared backups $s$$\qquad\ \ $}
					\State $s=\max_{i\in SNeigh(n_{host})}(m_{ij})$
					\State $\!\!\!\!$ \Comment{Allocate backups and update $u_{n_{host}}$ $\qquad\qquad\ \ $}
				  \State Allocate ($s$ backups, VNF type $j$, host $n_{host}$) 
					\State $BNodes(j) \gets BNodes(j) \cup SNeigh(n_{host})$
				\Until{$SNeigh(n)=\O\ \ \forall n \in N$}
		\EndFor
\end{algorithmic}
\end{algorithm}

\textbf{$\bullet$ Algorithm BS-Push:} 
in this algorithm, we are adopting an~approach different from the first one. For a particular physical node, our goal is maximize the number of~its~source neighbors that are using it (i.e., the physical node) to~host their VNF backups (unlike~BS-Pull that maximizes the~number of VNFs that are backed~up by the physical node but~not~the~number of~source neighbors using it).

As~shown~in~Algorithm~2, similar to BS-Pull, BS-Push computes the set of source neighbors for all the physical nodes (Line~8). However, the algorithm selects the node $n_{host}$ that  has the highest number of source neighbors in~order~to~host the~VNF backups of  all these neighbors (Line~10). The backup resources are then allocated and associated to all  neighbors of~the~selected node (Line~14). 
The operation is then repeated until there are no more source neighbors for all physical nodes. 
Finally, the whole process is applied again for each of~the~VNF~types.

\section{Simulation and Results}\label{Section:Simulations}
In this section, we compare the performance of the proposed algorithms with the optimal solution provided by CPLEX in~terms of~total number of backups and the execution time. 
To~do~so, we~implemented the algorithms~in~C and~simulated the~physical infrastructure and~the~service chain embedding. We have considered a~network with~24~physical nodes with different computing capacities randomly generated from 20 to 50 virtual machines. For simplicity, we assume that all virtual machines have the~same~resource capacities and~that~a~single~VNF is~hoted by a single virtual machine. Furthermore, the physical nodes are connected through 55 physical links that were randomly generated. We assume the~embedding of service chains (i.e.,~VNFs and virtual links) is~already carried out by an~existing VNF placement algorithm. In our experiments, we~used the resource allocation algorithm for service chains that~was~proposed~by~Racheg~et~al.~\cite{r10}. 

\begin{figure}[h!]
	\centering
	\includegraphics[scale=0.3]{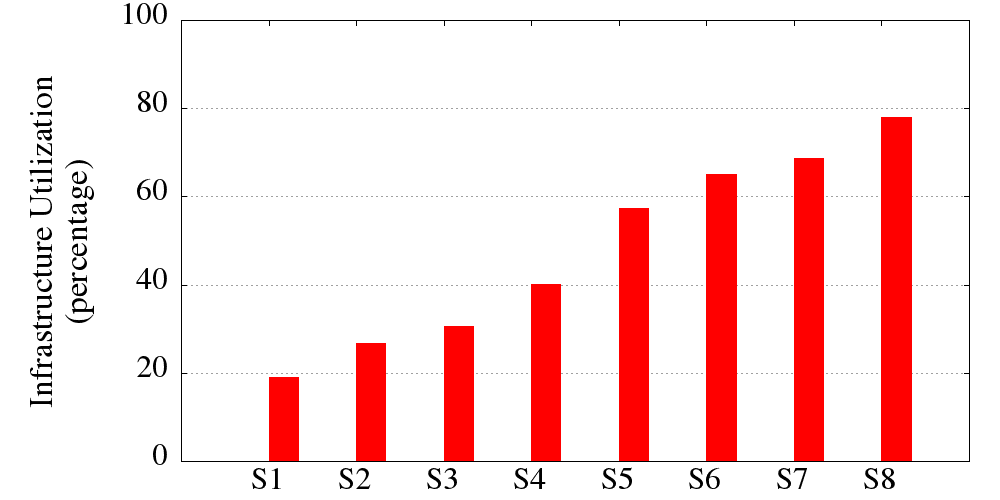}
	\caption{Studied scenarios with different infrastructure utilization.}
	\label{fig:InfraUtil}
\end{figure}

We considered 8 different embedding scenarios where the~utilization of the infrastructure has been gradually increased as~shown in~Figure~\ref{fig:InfraUtil}. We can see in the figure that we have low-utilization scenarios (i.e.,~s1 to s4) where utilization is~less~than~50\% and~also~high-utilization scenarios (e.g.,~s5 to s8) where utilization is~higher~than~50\%. 

The objective of the experiments is to compare the~number of~backups and the~number of~VNFs left without backup provided by our algorithms with the one provided by CPLEX.

\begin{figure}[h!]
	\centering
	\includegraphics[scale=0.3]{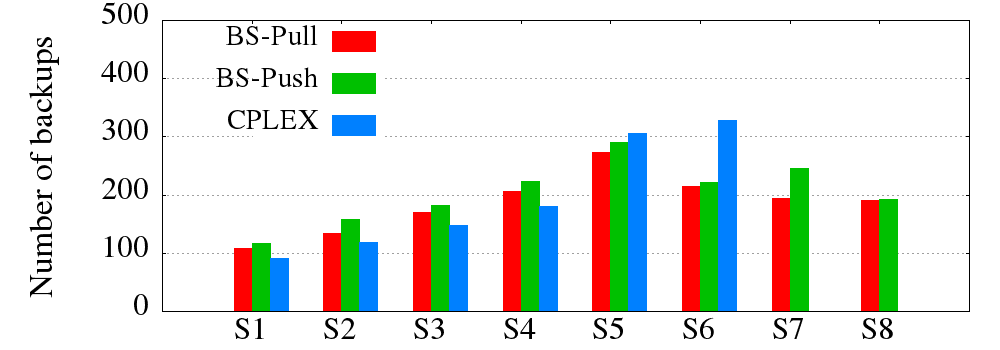}
	\caption{Total number of provisioned backups.}
	\label{fig:NumberBackups}
\end{figure}
\begin{figure}[h!]
	\centering
	\includegraphics[scale=0.3]{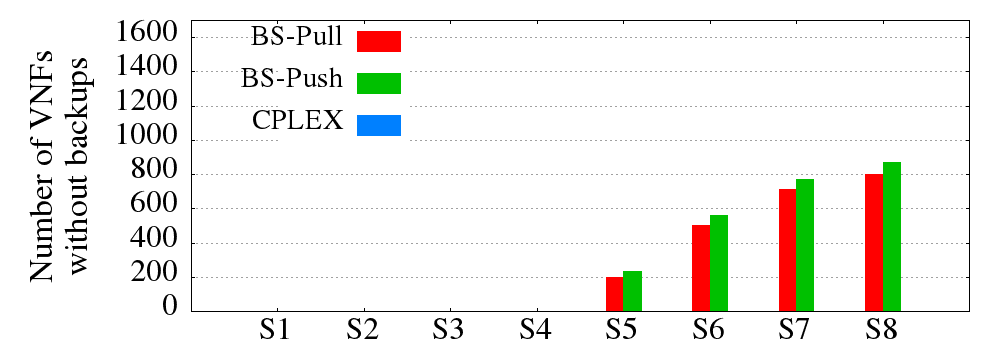}
	\caption{Number of VNFs without backup provisions.}
	\label{fig:NumberVNFBackups}
\end{figure}
\subsection{Number of backups}
Figure \ref{fig:NumberBackups} compares the total number of backups found with~the~proposed heuristic algorithms with~the~optimal solution produced by~CPLEX for each scenario.
We~can~see~that~for low-utilization scenarios (i.e.,~S1--S4), the~two~heuristics provide a slightly higher number of~backups compared to the optimal solution provided by~CPLEX in~low utilization scenarios, indicating that their solutions are~not~far from the optimal ones. In addition, we~notice~that~BS-Pull generally provides lower numbers of~backups compared~to~PS-Push.

For high-utilization scenarios (S5--S8), it becomes harder to~find even~a~feasible solution, i.e., a solution that~ensures that~all VNFs in the infrastructures have backups.
We~can~see~that,~for S5 and S6, only CPLEX could find a~solution (which~is~optimal) whereas the two heuristics do~not~ensure that there are backups for all VNFs as~depicted~in~Figure~\ref{fig:NumberVNFBackups}, which shows that many VNFs are~left without~backups using the~heuristics when utilization is~high.

Figure \ref{fig:NumberBackups} also shows that for scenarios S7 and S8 that~have high utilization (above 70\%), CPLEX does not find an optimal solution that allow to provide backups for~all~VNFs. This~is~because there is no enough free resources in~the~infrastructure to provision all~the~required backups. The~heuristics in this case still provide a solution even though several VNFs are left without backups as~shown in~Figure~\ref{fig:NumberVNFBackups}.

\subsection{Execution Time}
\begin{figure}[htb]
	\centering
	\includegraphics[scale=0.3]{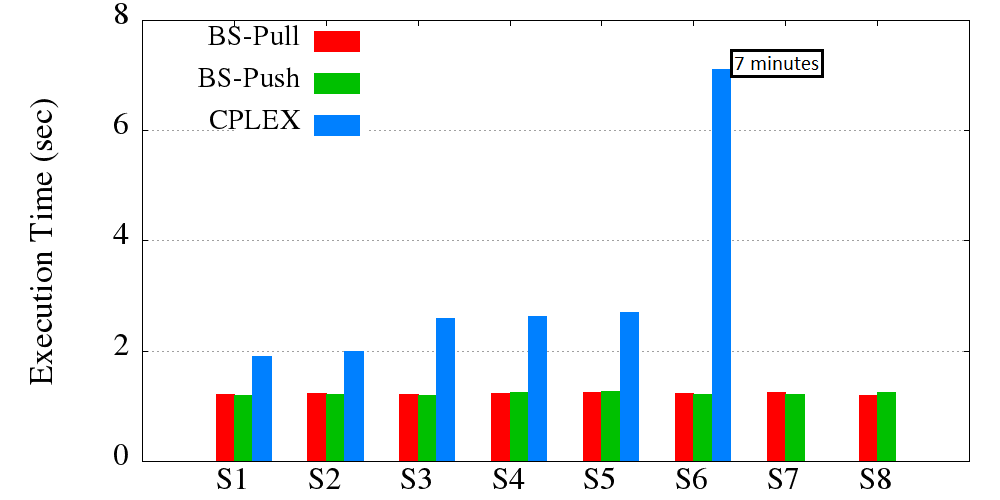}
	\caption{Execution time of the different algorithms.}
	\label{fig:ExecutionTime}
\end{figure}

Figure \ref{fig:ExecutionTime} depicts the execution time of~the~two~proposed algorithms compared to that of CPLEX for each of~the~8~studied~scenarios. 
The execution time for CPLEX goes from 2s to 7min (S6) as the infrastructure utilization increases. This is because the number of variables in the ILP increase (e.g., the number of nodes and that of VNFs) and makes the problem harder to solve because of the large research space. 
It is also clear from the figure that the two algorithms' execution times does not significantly change as~the~utilization of~the~infrastructure increases. We note, again, that beyond the sixth scenario (i.e.,~for S7--8), no~optimal solution could be found due~to~the~unavailability of~free~resources in~the~infrastructure.

\subsection{Synchronization Cost}

\begin{figure}[!h]
	\centering
	\includegraphics[scale=0.3]{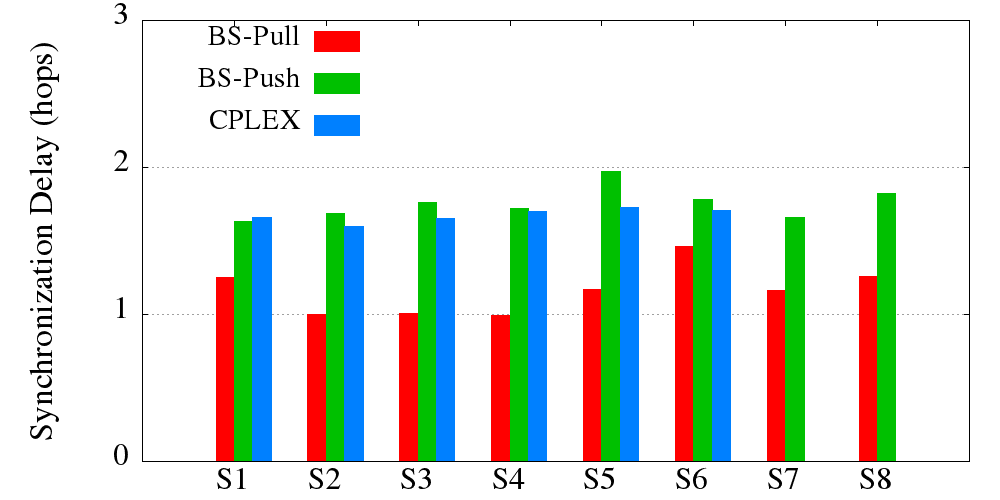}
	\caption{Synchronization distance of the different algorithms.}
	\label{fig:SyncDelay}
\end{figure}

Figure \ref{fig:SyncDelay} shows the average number of hops between an~embedded VNF and its backup for the different solutions and across the studied scenarios. The number of hops provides some insight about the potential synchronization cost in~terms of delay and bandwidth. The results show that~for~the~two~algorithms as~well~as~CPLEX, the~number of~hops is~below the maximal number of~hops $d_{max}$ specified as~input to~all~solutions (i.e.,~$d_{max}$~is~equal to~2~in~our~experiments). 

\section{Conclusion}\label{Section:Conclusion}
In this paper, we addressed one of the uprising challenges faced by the infrastructure providers: the survivability of~the~service chains against node failures. We hence proposed a novel solution that provision the minimal number of~shared backup VNFs that minimizes the amount of resources allocated for the backup.

We hence formulated the problem as an ILP and~then~proposed two heuristic algorithms to~solve the~problem for~large-scale scenarios. Through extensive simulations, we demonstrated that our algorithms provide solutions that are close to the optimal one provided by~CPLEX while they reduce the execution time considerably.

As future work, we aim to further optimize both algorithms in order to reduce their complexity. We~also~plan~to~extend~this~work to take into consideration multiple node failures.

\bibliographystyle{ieeetr}
{\balance 
\bibliography{biblio}
}

\end{document}